\documentclass[
twocolumn,
twoside,
letterpaper
]{IEEEtran}

\usepackage{amsmath,amssymb,amsfonts}
\usepackage{mathrsfs}
\usepackage{graphicx}

\usepackage[colorlinks,citecolor=red,urlcolor=blue,bookmarks=false,hypertexnames=true]{hyperref}

\usepackage{color}
\definecolor{brickred}{rgb}{0.8, 0.25, 0.33}

\begin{document}
\title{Measurable Improvement in Multi-Qubit Readout Using a Kinetic Inductance Traveling Wave Parametric Amplifier}

\author{M.A.~Castellanos-Beltran, 
        L.~Howe, 
        A.~Giachero,
        M.R.~Vissers,
        D.~Labranca,  
        J.N.~Ullom,
        P.F.~Hopkins
\	   
\thanks{M.A.~Castellanos-Beltran and P.F.~Hopkins  
        are with
        RF Technology Division, National Institute of Standards and Technology, Boulder Colorado 80305, USA
        }
\thanks{L.~Howe, M.R.~Vissers, and J.N.~Ullom
       are with 
       Quantum Sensors Division, National Institute of Standards and Technology, Boulder Colorado 80305, USA
       and also with
       Department of Physics, University of Colorado, Boulder Colorado 80309, USA
       }
\thanks{A.~Giachero 
       is with 
       Quantum Sensors Division, National Institute of Standards and Technology, Boulder Colorado 80305, USA, 
       also with
       Department of Physics, University of Colorado, Boulder Colorado 80309, USA
       and also with
       Department of Physics, University of Milano-Bicocca, 20126, Milan, Italy (e-mail: andrea.giachero@colorado.edu)
       }
\thanks{A.~Labranca   
        is with
        RF Technology Division, National Institute of Standards and Technology, Boulder Colorado 80305, USA
        and also with 
        Department of Physics, University of Milano-Bicocca, 20126, Milan, Italy
        }
}


\maketitle

\begin{abstract}


Increasing the size and complexity of quantum information systems requires highly-multiplexed readout architectures, as well as amplifier chains operating near the quantum limit (QL) of added noise. 
While documented prior efforts in KI-TWPA integration in quantum systems are scarce, in this work we demonstrate integration of a KI-TWPA with a multiplexed-qubit device. To quantify the system noise improvement we perform an ac Stark shift calibration to precisely determine noise power levels on-chip (at each cavity's reference plane) and the total system gain. We then characterize the qubit state measurement fidelity and the corresponding signal-to-noise ratio (SNR).

To conduct the most faithful measurement of the benefits offered by the KI-TWPA we perform these measurements for readout chains where the high electron mobility transistor (HEMT) amplifier is the first-stage amplifier (FSA) -- with none of the external hardware required to operate the KI-TWPA -- and with the KI-TWPA as the FSA. While some readout cavities fall outside the KI-TWPA bandwidth, for those inside the bandwidth we demonstrate a maximum improvement in the state measurement SNR by a factor of 1.45, and increase the fidelity from 96.2\% to 97.8\%. These measurements demonstrate a system noise below 5~quanta \textit{referenced on-chip} and we bound the KI-TWPA excess noise to be below 4~quanta for the six cavities inside its bandwidth. These results show a promising path forward for realizing quantum-limited readout chains in large qubit systems using a single parametric amplifier.

\end{abstract}


%
\IEEEpeerreviewmaketitle

\section{Introduction}\label{sec:intro}

Superconducting parametric amplifiers are critical components for the precise readout of qubits~\cite{Aumentado2020}. To achieve fast, high-fidelity readout~\cite{Jeffrey2014,Walter2017} it is paramount to amplify weak signals without introducing excessive noise, i.e the first-stage amplifier (FSA) must operate very close to the quantum limit (QL) of added noise~\cite{Clerk2010}. Josephson parametric amplifiers (JPAs)~\cite{Yurke1989,Siddiqi2004} have been a mainstay solution due to their low-noise performance~\cite{Yamamoto2008,Castellanos-Beltran2008}. However, as the number of superconducting qubits increases the need for simultaneous readout~\cite{Arute2019,Bravyi2022} of a large number of devices severely hampers feasible scaling using JPAs. Notably, JPAs are very narrowband and these Josephson-junction-based amplifiers typically have an input dynamic range below -100~dBm \cite{qiu2023broadband, ranadive2022kerr}. Impedance transformers and parallel distribution of the amplification can be implmented to modestly increase bandwidth and dynamic range, however, this comes at the cost of significant added complexity \cite{kaufman2023josephson}. An alternate technology for achieving broadband, quantum-limited readout is that of the Traveling Wave Parametric Amplifier (TWPA) \cite{Esposito2021}. Superconducting TWPAs exploit the non-linearity of inductive elements embedded in a long transmission line (TL) to amplify weak microwave signals with minimal added noise \cite{Cullen1960}. TWPAs based on Josephson junctions~\cite{Macklin2015} (JTWPAs) are a potential solution for multiplexed qubit readout. 

Another solution is a TWPA based on the non-linearity of kinetic inductance in disordered materials~\cite{Erickson2017,HoEom2012} (KI-TWPAs). Although proof-of-principle demonstrations in qubit readout exist \cite{Ranzani2018}, KI-TWPAs have not enjoyed the same extensive integration as Josephson-junction based parametric amplifiers. This is contrary to the most attractive features of KI-TWPAs, relative to JPAs and JTWPAs: they are simpler to fabricate~\cite{Giachero2023} and can provide a dynamic range orders of magnitude higher \cite{vissers2016low, faramarzi20244a4to8}. Further, they have demonstrated power gains above 15~dB over 3~GHz of bandwidth~\cite{Malnou2021, Shu2021}, they can operate at higher temperatures~\cite{Malnou2022}, and are resilient to high magnetic fields\cite{Vaartjes2024,Frasca2024} -- reducing the amount of shielding required.

Indeed, one of the primary obstacles in KI-TWPA adoption lies in their historically high pump power requirements~\cite{vissers2016low, ho2012wideband, Malnou2021}, making thermalization of the pump injection line impossible at a level compatible with high fidelity qubit readout. However, by reducing the high kinetic inductance film thickness from 20~nm to 10~nm we have recently demonstrated KI-TWPAs operating with an over tenfold reduction in the pump power \cite{Giachero2024, Malnou2021}. These devices significantly enhance the feasibility of qubit-KI-TWPA integration as (\textit{i}) now the pump line may be appropriately thermalized so as not to affect qubit coherence, and (\textit{ii}) heating of the mixing chamber stage is now negligible. 

In this study, we integrate one such KI-TWPA (requiring a -43~dBm pump power at the package input) as the FSA in an eight-qubit readout chain and demonstrate an improvement in the qubit state measurement fidelity, its associated signal-to-ratio (SNR), and system noise relative to readout performed with a HEMT as the FSA. 

\begin{figure}[!t]
\centering \includegraphics[width=0.45\textwidth,clip]{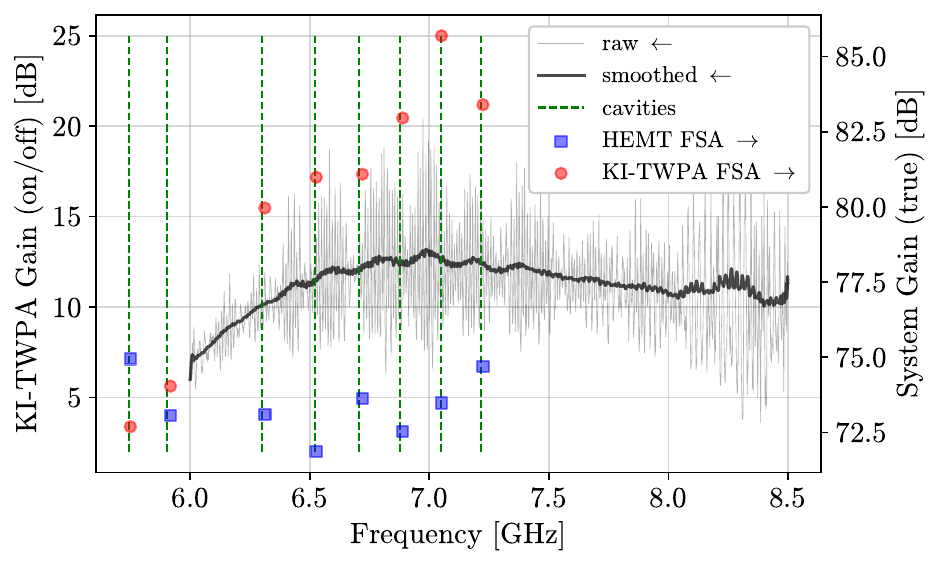}    
\caption{(\textit{left-axis}) Measured gain (on/off) of the KI-TWPA used in this work. The gain ripple periodicity is of order 25~MHz and is thus much wider than the readout cavities. (\textit{right-axis}) System gain for the two experimental configurations extracted using the ac Stark calibration. This demonstrates that the true gain added to the readout chain by incorporating the KI-TWPA is not the on/off gain, rather it is the on/off gain reduced by the KI-TWPA's internal loss -- which is typically $0.5\text{--}1$~dB/GHz. This is particularly relevant for the two qubits with the lowest cavity frequencies. These resonators fall far enough outside the KI-TWPA gain bandwidth that the KI-TWPA acts as a lossy transmission line (attenuator) and thus increases the system noise temperature for these two qubits.}
\label{fig:gain}
\end{figure}

The KI-TWPA used in this work is similar to that of~\cite{Giachero2024} where the transmission line is a stub-loaded inverted microstrip ~\cite{Chaudhuri2017,Malnou2021,Shu2021}. A representative gain (on/off) profile is shown in Fig.~\ref{fig:gain}. The signal layer is patterned from a $t=10$\,nm thick NbTiN film with a kinetic inductance of $L_0=35$\,pH/$\square$. The dielectric is $d=100$\,nm of amorphous Si ($\alpha$-Si) with a relative permittivity of 9.1, and the (top) ground plane is made of 100\,nm thick niobium (negligible $L_0$). The width of the microstrip and of the stubs is $1\,\mu$m. The spacing between two adjacent stubs is $1\,\mu\text{m}$. Since $Z_0 = \sqrt{\mathcal{L}/\mathcal{C}}$, the line characteristic impedance is adjusted by tuning the stub length. The transmission line dispersion is engineered to create broadband phase matching for first-order parametric processes and amplification -- while suppressing higher order processes such as third harmonic generation. Electromagnetic simulations provided a stub length of $11\,\mu\text{m}$ ($3.5\,\mu\text{m}$) to create $Z_0 = 50\,\Omega$ ($Z_0 = 80\,\Omega$) cells to perform this necessary dispersion engineering. These parameters mean that these devices have an average inductance (capacitance) per unit length of $\mathcal{L} = 35$~pH/$\mu$m ($\mathcal{C} = 9.3$~fF/$\mu$m). This is achieved with a series of 1200 \textit{supercells}, each consisting of 30 \textit{unloaded cells} ($50\,\Omega$) and 4 \textit{loaded cells} ($80\,\Omega$). The total transmission line length is 8.3\,cm (41276 cells, each $2\,\mu$m long). 


\section{\label{sec:exp_setup}Experimental Setup}
\begin{figure*}[t!]
    \centering\includegraphics[width = .7\textwidth]{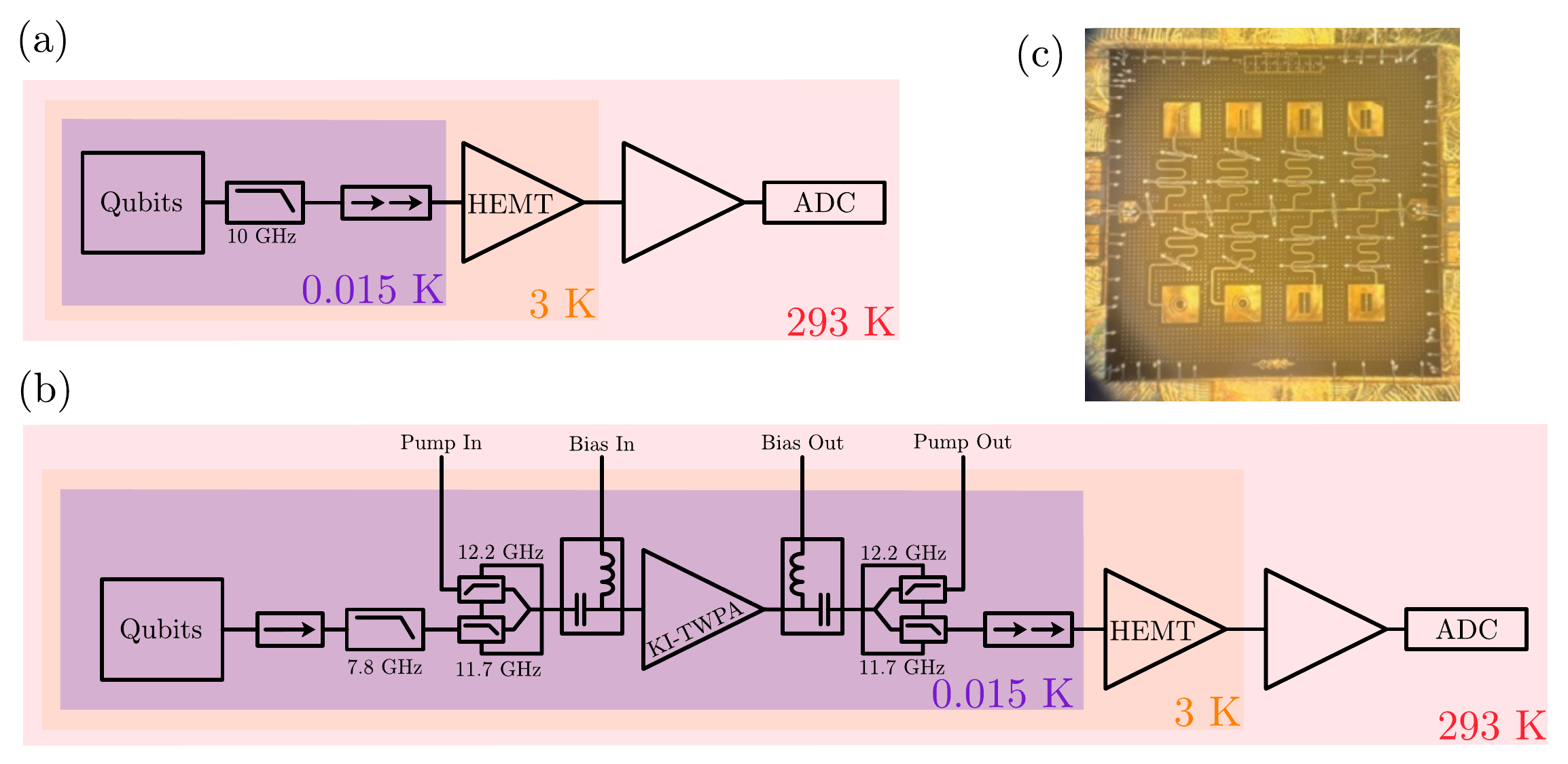}
    \caption{Schematics of the experimental setups used in this work. In both cases a lowpass filter (10~GHz in (a), and 7.8~GHz in (b)) and double-stage isolator prevent the HEMT noise from affecting qubit coherence. \textbf{(a)} Setup with the HEMT as the first-stage amplifier. \textbf{(b)} Setup using the KI-TWPA as the first-stage amplifier. We apply the pump to the highpass port of the input diplexer and then route the remnant pump power up to room temperature using the output diplexer to prevent saturation of the HEMT. The dc bias is applied via bias tees using a voltage source in a source/sink configuration with 10~k$\Omega$ current-limiting resistors at room temperature. To minimize the impact on the qubits from amplified vacuum noise which is reflected at the KI-TWPA output we place a single-stage isolator between the qubit and the KI-TWPA. \textbf{(c)} Packaged qubit chip.}
    \label{fig:schematic}
\end{figure*}

Fig.~\ref{fig:schematic} shows the dilution refrigerator experimental schematic for the two configurations. Our goal is to directly compare the qubit readout fidelity and system noise in two cases: (a) HEMT as the FSA, and (b) KI-TWPA as the FSA with amplification on. Often a TWPA-on versus -off comparison is made to quote performance but this results in an over-estimation of the TWPA gain by an amount equal to the TWPA's internal loss; plus loss from all the peripheral components required to operate the TWPA (typically \mbox{1--2~dB}). For many TWPAs the internal loss is non-negligible \cite{Macklin2015, simbierowicz2021characterizing, qiu2023broadband, ranadive2022kerr, howe2025compact} and in our case it is approximately $0.5\text{--}1$~dB/GHz, i.e. with the TWPA installed but not pumped, this is akin to deliberately placing a large attenuator between the DUT and FSA. 
A more accurate comparison of the system performance  is shown in the setups in Fig.~\ref{fig:schematic}, where when the HEMT is the FSA we remove the KI-TWPA and all its associated external hardware.

The qubit chip used in this work, shown in Fig.~\ref{fig:schematic}(c), consists of eight fixed frequency transmon qubits coupled to a common feedline through their readout cavities. Typically this device is used to diagnose qubit relaxation time $T_1$ vs various fabrication parameters; thus all qubits are necessarily very weakly coupled to the feedline. This does not facilitate fast readout protocols~\cite{Walter2017} and such an architecture does not fully leverage the benefit offered by a lower system noise when using the KI-TWPA. In the future a qubit device more optimized for fast and high fidelity readout will be used.

\section{Calibration Procedure}


To measure the system noise at the cavity reference plane we use each qubit as a power meter using the ac Stark effect \cite{Schuster2005}. Qubit spectroscopy as a function of the readout power (i.e. cavity photon number) can be used to precisely calibrate the incoming power at the reference plane of the cavity. As the cavity photon number increases, the qubit frequency is Stark-shifted to lower frequency by an amount directly related to the qubit-cavity coupling~\cite{Schuster2005,Vijay2012}. This calibration is performed for each of the eight qubits in the case of the HEMT as the FSA and then recalibrated with the KI-TWPA as the FSA.


\begin{table}
\caption{Cavity-qubit parameters required for calibration}
\begin{center}

\label{table:1}
\begin{tabular}{||c c c c c||} 
 \hline
  & $f_{res} $ (GHz) & $ Q_{c} $  & $\chi  $ (kHz) &  $f_{q} $ (GHz)  \\ [0.5ex] 
 \hline\hline
 1 & 7.218  &$7136 \pm   46$  & $135 \pm 5$ &   4.730\\ 
 \hline
 2 & 7.048  &  $4216 \pm 27 $  & $140 \pm 12 $ & 4.583  \\
 \hline
 3 & 6.879  & $7842\pm 90$   &  $159 \pm 8 $   & 4.553 \\
 \hline
 4 & 6.707  &  $6603 \pm 57$ &   $95 \pm 7$ &  3.399\\
 \hline
 5 &   6.522 & $7843 \pm 64$    &$153 \pm 8$ & 4.288 \\
 \hline
 6 &  6.299  & $9642 \pm 87$    & $143 \pm 12$ & 4.066 \\
 \hline
 7 & 5.903  & $5814 \pm 35 $    & $156 \pm 15$ &  4.015\\
 \hline
 8 &  5.745  & $ 11290\pm 90 $    & $265 \pm 14$ & 4.411   \\
 \hline
\end{tabular}

\end{center}

\end{table}

\subsection{Cavity-Qubit parameters extraction}

\begin{figure}
    \centering\includegraphics[width = .45\textwidth]{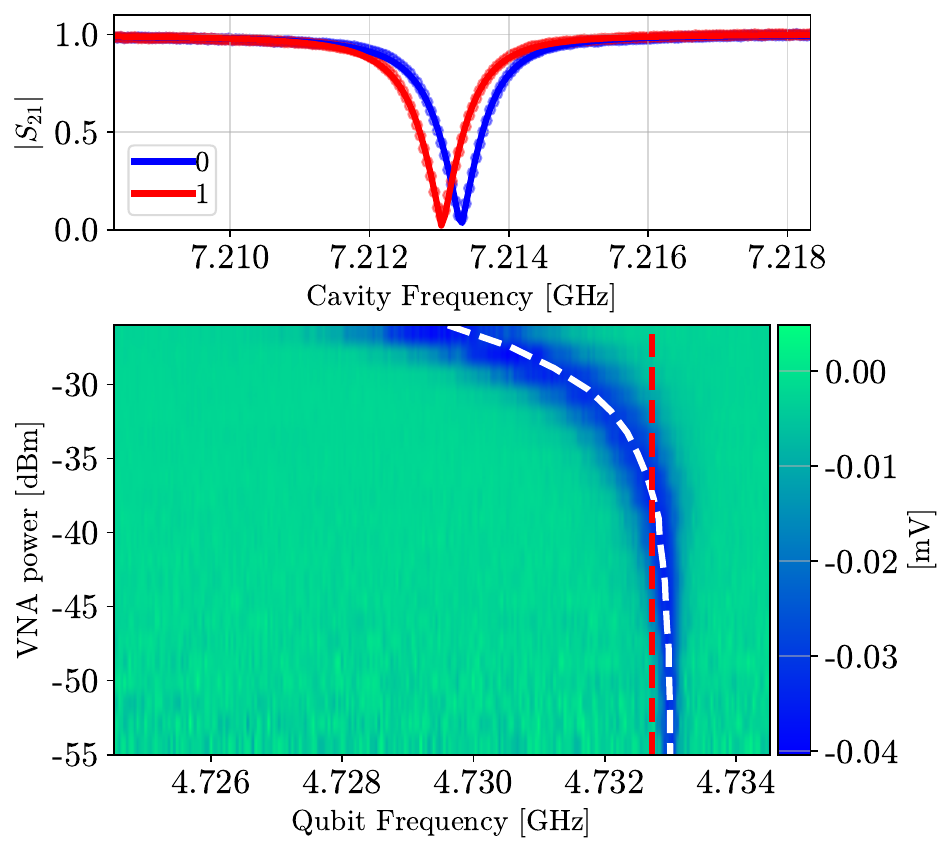}
    \caption{(a) Extraction of the dispersive shift $2\chi$ for a single resonator. (b) Qubit spectroscopy in the presence of readout photons as a function of readout tone power. The qubit transition frequency shifts, in response to the ac Stark shift, and broadens, due to the increased cavity occupation. The white dashed line is the extracted qubit frequency for at each readout power. The red dashed line is the intrinsic qubit frequency shifted by $2 \chi$. When the two dashed lines meet, the mean cavity occupation is $\langle N_{c}\rangle=1$.}
    \label{fig:acstark}
\end{figure}

We first extract  the resonator parameters (resonance frequency, internal, and external quality factors) that are necessary for the estimation of the cavity photon number occupation~\cite{Bruno2015,McRae2020}. Afterwards, we extract the qubit parameters: both the transition frequency, $\omega_q$, as well as the  dispersive coupling strength between the resonator and the qubit, $\chi$. The latter is extracted by measuring the cavity response versus drive frequency with the qubit in the 1 and 0 state. Due to the high relaxation times of these qubits, the cavity response is approximately the same other than shifted by $2\chi$~\cite{Bianchetti2009}. Fig.~\ref{fig:acstark}(a) shows an example of the dispersive shift extraction for a single qubit and cavity. Parameters for all eight qubits are shown in Table~\ref{table:1}.

Next, we perform the ac Stark shift power calibration via qubit (two-tone) spectroscopy. The cavity is driven on resonance and the qubit drive frequency is swept. An example is shown in Fig.~\ref{fig:acstark}(b). When the qubit frequency shifts by $2\chi$ (red dashed line), the cavity is in a coherent state with an average photon occupation number of 1 quanta ($\langle N_{c}\rangle =1$). The readout tone power at the cavity reference plane is now known via $\langle N_c \rangle$ and the relation 
\begin{equation}
P_{\text{cav}}=\frac{Q_c}{Q^2}\frac{\langle N_{c}\rangle\hbar \omega^2}{2}\simeq \frac{1}{Q}\frac{\langle N_{c}\rangle\hbar \omega^2}{2},
\label{eq:cav_power_acStark}
\end{equation}
where $Q_c$ is the coupling quality factor, $Q$ is the total quality factor and $P_{\text{cav}}$ is the power (in Watts) at the cavity input port~\cite{Bruno2015}. Eq.~(\ref{eq:cav_power_acStark}) allows us to easily calibrate the signal-generator-qubit attenuation and thus accurately know the on-chip power level at each cavity frequency. The fitted losses in the resonators are very low compared to the designed $Q_c$ (fitted internal quality factor $Q_\text{int}\ge~500\times Q_c$). Given the low accuracy in estimating $Q_\text{int}$ due to this large asymmetry \cite{McRae2020}, we assume that $Q=Q_c$ when calculating the on-chip power level. 

\subsection{Noise Calibration}

\begin{figure}
    \centering\includegraphics[width = .45\textwidth]{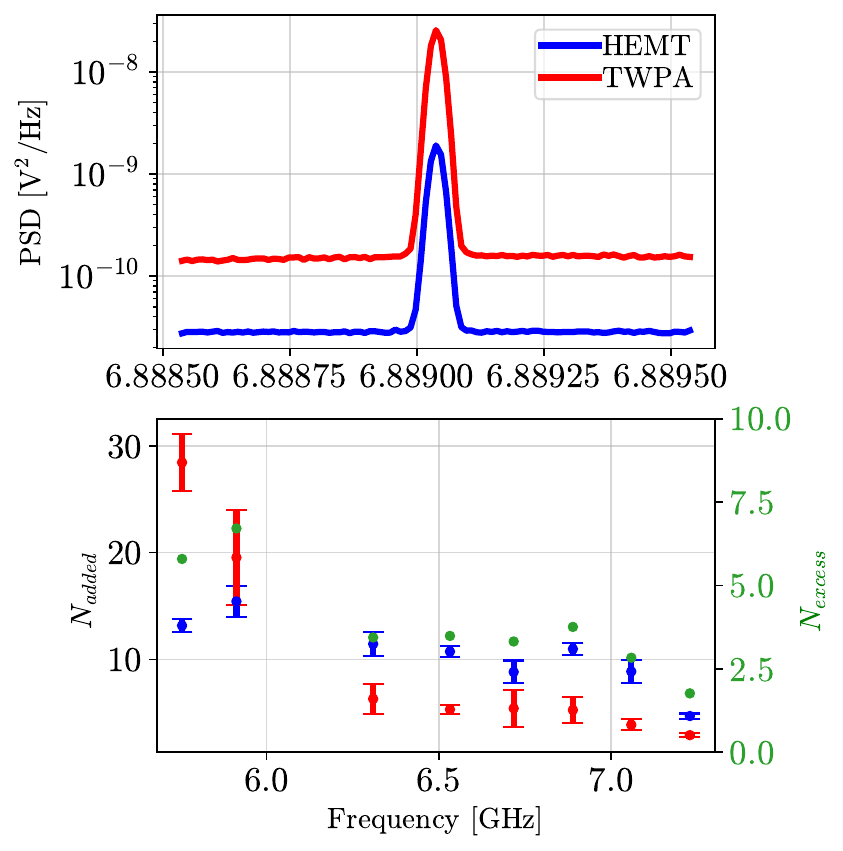}
    \caption {(a) Example power spectrum in the vicinity of a single readout cavity (cavity \#3) using an ac-Stark-calibrated probe tone. (b) Extracted added noise for each of the HEMT and KI-TWPA as the FSA cases. Using Eq.~(\ref{eq:noise}) and the known tone power on-chip we extract the system noise detuned by 10~MHz from each of the eight resonators. Note these values are for the system noise with an on-chip reference plane -- i.e. on the other side of the lossy isolator, filters, diplexer, and bias tee -- which easily increase the system noise by two or more quanta relative to the reference being at the KI-TWPA input. The errorbars are dominated by the uncertainty in the cavity-qubit parameter fits and not the measurement error of the signal and noise. The right axis shows the estimated excess noise above the QL of 1/2 a  quanta. Error bars were not plotted for clarity but are similar in size to the red data points.}
    \label{fig:Tn}
\end{figure}

Having calibrated the attenuation from room temperature to the chip, it is possible to measure the system (total) gain of the amplification chain $G_{sys}$ when the TWPA is and is not present. We extract the system gain by sending the now-calibrated input signal through the qubit chip, and then measure its output power at room temperature with a spectrum analyzer (Fig. \ref{fig:Tn}a). With the system gain we can then obtain the noise floor power per unit Hertz (power spectral density) at the reference plane of each cavity (in photon-normalized units)
\begin{equation}
N_{added}=\left( k_B \left[ \frac{ P^{SA}_{noise}}{G_{sys}k_B BW} \right] - \frac{\hbar \omega}{2} \right) \frac{1}{\hbar \omega},
\label{eq:noise}
\end{equation}
where $BW=30$~kHz is the measurement bandwidth, $k_B$ and $\hbar$ are the Boltzmann and reduced Planck constants, respectively, $\omega$ is the cavity resonance frequency, and the expression inside the square brackets is the system noise temperature $T_N$~\cite{Qiu2023}. The results of these measurements are shown in Fig.~\ref{fig:Tn}(b) for the two cases of (a) HEMT and (b) KI-TWPA as the FSA -- showing a small but measurable improvement in the system noise with the KI-TWPA as the FSA.

\begin{figure}
    \centering\includegraphics[width = .45\textwidth]{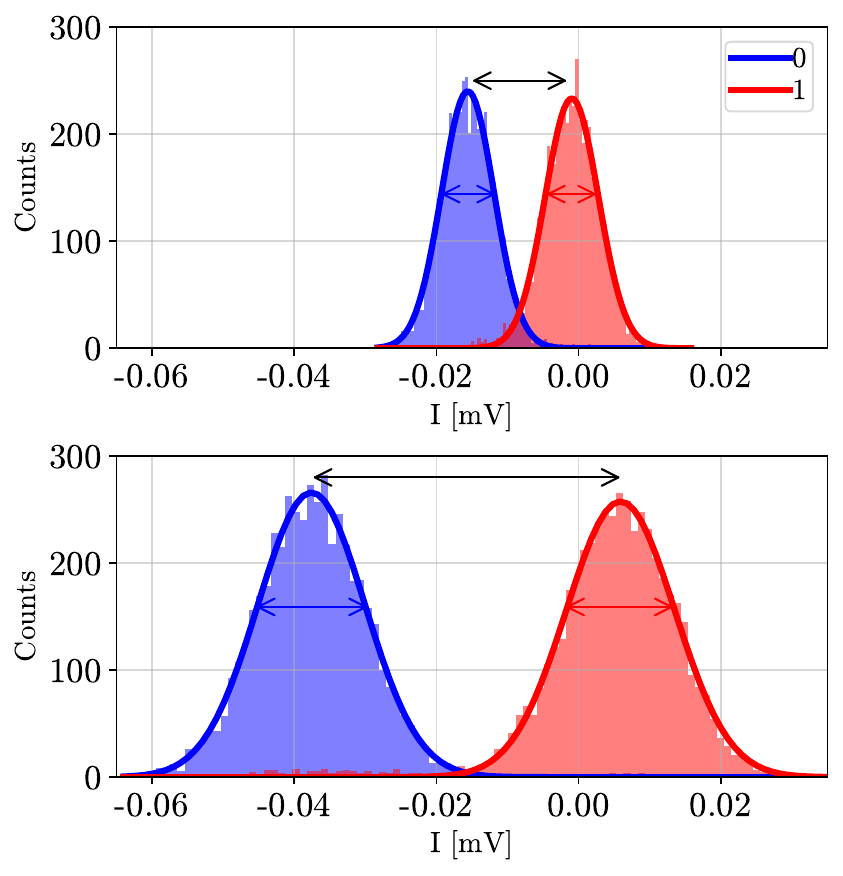}
    \caption{SNR improvement comparison between the (a) HEMT- and (b) KI-TWPA-as-FSA dispersive readout of the qubit states. The signals in both cases are indicated by the black arrows and the standard deviations of the Gaussian fits by the blue ($|0\rangle$) and red ($|1\rangle)$) arrows.  }
    \label{fig:measfid}
\end{figure}

The observed improvement in system noise is also reflected in an improvement of the qubit state measurement fidelity. This is shown in Fig.~\ref{fig:measfid}, for a single qubit, where the result of single shot measurements are shown for both the (a) HEMT and the (b) KI-TWPA as the FSA. The measurement fidelity defined as 
\begin{equation}
\mathcal{F}= 1-(P(0|1)+P(1|0))/2
\end{equation}
improves from 96.2\% to 97.8\%~\cite{Magesan2015}. An alternative metric to show the improvement provided by the KI-TWPA is to measure the SNR of the state measurement fidelity, defined as the distance between the two Gaussian histograms scaled by their standard deviations~\cite{Lecocq2021}. The single-shot measurement signals after an integration time $\tau = 1~\mu$s, and after rotating into a single ($I$) quadrature, shows two normal distributions with means $I_{0,1}$ and standard deviations $\sigma_{0,1}$. The ratio of the distance between the histogram means (black arrows) $\mu= |I_0 - I_1|$ to the total standard deviation $\sigma_T=\sqrt{\sigma_1^2+\sigma_0^2}$ gives the SNR:
\begin{equation}
SNR=\frac{\mu}{\sigma_T}.
\end{equation} 
We observe an improvement in SNR by a factor of 1.45, similar to the observed improvement of the amplification chain system noise: $\sqrt{T_{N,HEMT}/T_{N,TWPA}}$.

\section{Conclusion}
We have demonstrated the efficacy of performing multi-qubit readout using a Kinetic Inductance Traveling Wave Parametric Amplifier (KI-TWPA),  and determined the noise spectral density on-chip using the ac Stark effect as an absolute power calibration. 
We have directly compared the system noise and qubit state measurement fidelity by conducting two separate cooldowns with different first-stage amplifier (FSA) configurations as shown in Fig. \ref{fig:schematic}. By performing separate cooldowns, we  account for both the increased hardware overhead 
and the internal loss of the KI-TWPA itself by only comparing the HEMT-only versus KI-TWPA-on system performance. 
Our results demonstrate a small increase in qubit state measurement fidelity and moderate improvement in the signal-to-noise-ratio. The improvement was limited primarily by the maximum KI-TWPA gain, which was over 10~dB lower than the median performance of other dies on the same wafer. Future implementations will utilize higher gain KI-TWPAs, and even KI-TWPAs featuring on-chip integration of the required rf components (for bias and pump injection), to demonstrate even more substantial qubit readout performance improvements enables by a KI-TWPA as the FSA.

\section*{Acknowledgment}
This work is supported by the National Aeronautics and Space Administration (NASA) under Grant No. NNH18ZDA001N-APRA, the Department of Energy (DOE) Accelerator and Detector Research Program under Grant No. 89243020SSC000058, and DARTWARS, a project funded by the European Union’s H2020-MSCA under Grant No. 101027746. The work is also supported by the Italian National Quantum Science and Technology Institute through the PNRR MUR Project under Grant PE0000023-NQSTI. We thank D. Olaya at NIST, and M. Bal and S. Zhu at FermiLab’s Superconducting Quantum
Materials and Systems (SQMS) Center, for the design, fab, and packaging of the qubit chip.

\bibliographystyle{IEEEtran}
\bibliography{main_IEEE_format_ASC2024_new}

\end{document}